# An Objective Assessment of the Utility of a Driving Simulator for Low Mu Testing

**Romano R** [a], **Markkula, G** [a], **Boer E** [a], **Jamson H** [a], **Bean A** [b], **Tomlinson A** [a], **Horrobin A** [a], **Sadraei E** [a]

[a]University of Leeds, LS2 9JT, Leeds, United Kingdom
[b]Jaguar Land Rover, CV35 0RR, Gaydon, United Kingdom

***Abstract*** – *Driving simulators can be used to test vehicle designs earlier, prior to building physical prototypes. One area of particular interest is winter testing since testing is limited to specific times of year and specific regions in the world. To ensure that the simulator is fit for purpose, an objective assessment is required. In this study a simulator and real world comparison was performed with three simulator configurations (standard, no steering torque, no motion) to assess the ability of a utility triplet of analyses to be able to quantify the differences between the real world and the different simulator configurations. The results suggest that the utility triplet is effective in measuring the differences in simulator configurations and that the developed "Virtual Sweden" environment achieved rather good behavioural fidelity in the sense of preserving absolute levels of many measures of behaviour. The main limitation in the simulated environment seemed to be the poor match of the dynamic lateral friction limit on snow and ice when compared to the real world.*

## 1. Introduction

A frequent use of driving simulation by automotive companies is to support the vehicle design process. While the range of use cases is broad, with the driving simulator sometimes being used as a task loading device, one goal for automotive companies is to reduce physical tests by performing virtual testing. To achieve this goal it is expected that the test driver in the simulator must perform similarly to the real world. The Objective Motion Cueing Test (OMCT) from flight simulation is starting to be applied to ground vehicle simulators (Fischer et al., 2016) to assess how closely the motion cues in a simulator are to the vehicle accelerations generated by the vehicle dynamics model. While this is a useful start, the goal of the current research is to define and test a methodological approach: the utility triplet (Boer et al., 2014), to assess behavioural fidelity: the degree to which drivers behave the same way in the simulator as in the real vehicle.

Three alternative configurations of the University of Leeds Driving Simulator (UoLDS) were chosen (STD) a "standard" UoLDS configuration, with the full motion system as well as steering wheel torque feedback turned on, (-MOT) a configuration where the motion system was turned off, but the steering torques left on, and (-TRQ) the opposite, with the motion system on but steering torques off (Boer et al., 2014). This allows an investigation of the drivers' use of the three main types of perceptual cues that were considered relevant for the targeted tasks: visual, vestibular, and haptic steering cues. The driving performance in the simulator was compared with similar manoeuvres performed on the real proving ground.

### 1.1 Review of the Literature

Previous research into the impact of steering torque (Mohellebi et al., 2009; Toffin et al., 2003) found that the type of steering torque feedback had an impact on lane-keeping and that a complete absence of steering torque degraded lane-keeping performance possibly even to the point of drivers giving up.

Previous research into the effects of motion cueing is large. It is well documented that adding some form of motion cueing generally increases the subjectively perceived realism, compared to fixed-base simulation (Damveld et al., 2012). However, and more interestingly, large-scale simulator motion that is objectively closer to the actual vehicle motion in the given scenario is not always perceived as more realistic, and for demanding lateral tasks such as slaloms it has been shown that drivers often prefer a motion scaling of about 0.5 over scale factors closer to 1 (Jamson, 2010; Savona et al., 2014). One possible explanation could be that the larger scale factors make false cues, arising from motion cueing imperfections, more salient. However, Berthoz et al argue that drivers prefer the smaller scale factors even in cases where no obvious false cues are present, and instead suggest that vehicle control gets more demanding for drivers who are subjected to large accelerations, and that this is effectively what drivers are subjectively reporting on (Berthoz et al., 2013) . This type of interpretation, which suggests that drivers might seek to avoid large accelerations, is in line with findings of drivers adapting to speeds and driving trajectories that limit the experienced acceleration, when motion cues are added or increased (Correia Grácio et al., 2009; Siegler et al., 2001). For objective task performance, just as with subjective realism, it generally improves when motion cues are first added and this performance increase is typically associated with decreases in measures of control effort, such as steering wheel reversal rate or high-frequency steering power (Damveld et al., 2012; Feenstra et al., 2010; Repa et al., 1982).

One point to note regarding these findings on objective performance, is that the performance has often been defined in terms of deviation from a researcher-defined reference path, for which there may be no strong motivation. Therefore, if drivers respond to higher motion scaling by adapting to an acceleration-limiting slalom trajectory that still avoids all cones (Berthoz et al., 2013), it could be argued that although the deviation from some normative path has increased, objective task performance has not really deteriorated. Conversely, if adding more motion cues or even outright false cues (Jamson, 2010) improves objective performance according to a researcher-defined metric, this does not necessarily mean that the behaviour has become more similar to what it would be like in a real car.  Therefore an objective comparison between reality and the simulator is required; however, validation studies to date have focused mainly on routine driving (Boer et al., 2000; Engström et al., 2005; Klee et al., 1999; Klüver et al., 2016; Wang et al., 2010).

The present study assesses behavioural fidelity in near-limit, vehicle-development type driving tasks. One factor helping to make this possible is the limitation of the scope to low-friction manoeuvring, which keeps accelerations within more feasible ranges for a simulator motion system.  Denoual et al. studied  (Denoual et al., 2011) loss of adherence and found that drivers were able to discriminate between difference friction levels and that both steering and motion cues were important in identifying the loss of adherence. Others have performed validation of the vehicle models without the driver in the loop (Gil Gómez et al., 2017; Gomez et al., 2016).

## 1.2 Research Questions

The experiment was designed to answer the following research questions:

1. In the highest-specification UoLDS simulator configuration, how similar is driver behaviour in the winter testing tasks to behaviour in a real vehicle (behavioural fidelity)
2. How is behavioural fidelity in the simulated winter testing tasks affected by altering the vestibular and haptic cues that are available to the drivers (impact of perceptual cues)
3. How do the drivers' subjective impressions of simulator realism relate to objective measures of behavioural fidelity (subjective impressions)

# 2. Method

Ethical approval was grant by the University of Leeds prior to data collection (LTTRAN-017).  Driver behaviour and vehicle data were collected both on the Revi test tracks in Sweden and on a virtual replica ("Virtual Sweden") of this test environment in the University of Leeds Driving Simulator (UoLDS), across the different simulator configurations (motion on/off, steering torque on/off). Three different manoeuvring tasks were tested in both of these environments: a lane change across ice (LCT), a slalom (SLX), and a circular curve (CLV). The objective measures outlined in Table 1 below were complemented with subjective ratings of simulator realism.  The objective measures of aggregate performance, time series analysis, and driver model fitting were developed previously (Boer et al., 2014).  The driver model is fit open loop to predict the driver's gain and response delay. A desired path yaw rate error (DPYRE) model was used.  DPYRE calculates the yaw rate that, starting from the current vehicle position and heading, would make the vehicle's trajectory intersect the desired path after a preview time $T_P$. This is compared with the actual steering input from the driver to yield a gain K and a response delay $T_R$ as given in Eq. 1.

$$\dot{\delta}_{SW}(t) = -K \cdot \omega_{err}(t - T_R) \qquad (1)$$

The model was extended for the CLV manoeuvre to an intermittent DPYRE model (Markkula et al., 2018) as explained in the results section.

Table 1. Objective Measures

|     | Aggregate performance | Time series | Driver model |
|-----|----------------------|-------------|--------------|
| LCT | • Cones knocked over<br>• Speed variability<br>• Max lateral acceleration | • Initial speed<br>• Total lateral travel<br>• Steering wheel reversal rate (SWRR) 1 deg<br>• SWRR 10 deg | DPYRE model:<br>• Response delay $T_R$<br>• Steering gain $K$<br>• RMS error |
| SLX | • Cones knocked over<br>• Speed variability | • Average speed<br>• Average peak lateral amplitude<br>• Peak lateral amplitude variability<br>• SWRR 1 deg<br>• SWRR 10 deg | DPYRE model:<br>• Response delay $T_R$<br>• Steering gain $K$<br>• RMS error |
| CLV | • Speed variability<br>• Turning radius variability | • Average speed<br>• Average radius<br>• SWRR 1 deg<br>• SWRR 10 deg | Intermittent DPYRE model:<br>• Response delay $T_R$<br>• Steering gain $k$<br>• RMS error<br>• Average adjustment magnitude |

Eight drivers were involved in the study, all JLR employees with a range of experience winter testing vehicles. Among the eight drivers the prior experience of winter testing, before this visit to Sweden, ranged from one season (two drivers), to two seasons (two drivers), to 15-30 seasons (four drivers).

## 2.1 Driving Manoeuvres
### 2.1.1 LCT: Single Lane Change μ Transition with throttle control
This task was a wide, single lane-change, undertaken on packed snow and polished ice. Starting on packed snow, the participant driver approached a coned gate at fixed speed and gear before manoeuvring sharply to the right, across the ice, before straightening up on the packed snow on the opposite side to complete the lane-change successfully by passing through two subsequent gates. Participants were instructed to attempt to maintain 45 kph and to manage the lane change as smoothly as possible without hitting any cones. The cone layout for the lane-change across ice and the relative location of the section of polished ice are shown below in Figure 1.

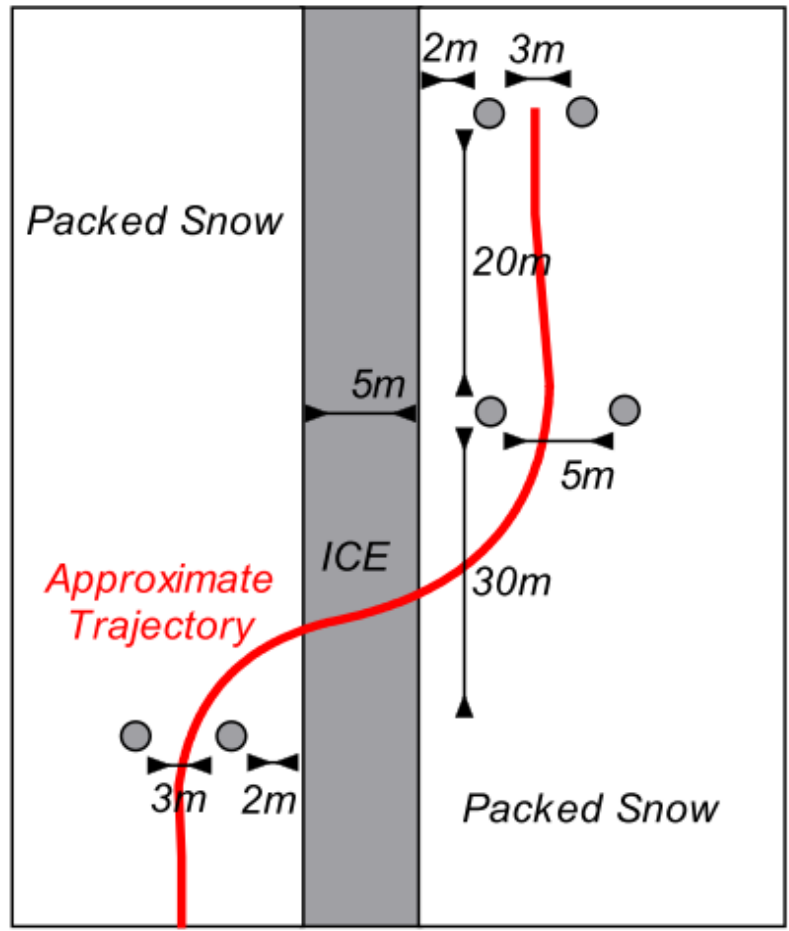

**Figure 1. Single Lane Change µ Transition**

### 2.1.2 SLX: Slalom
The slalom took place on packed snow. After the entry gate (two cones 5m apart), there were eight cones to negotiate, each spaced by 25m. The task involved rapidly steering the vehicle from its initial lane to a parallel lane around each cone without striking any. The slalom was performed at 45kph throughout in 2nd gear.

### 2.1.3 CLV: Curve Launch / Vmax combo
The curve launch was executed on packed snow. The driver was instructed to position the vehicle about half a car width from the inner edge of the inner radius of a small circle facing in the clockwise direction. The start location was marked by two adjacent cones. With manual transmission selected in second gear, the task was to accelerate from rest to the maximum controllable speed of the vehicle (Vmax) within the first lap of the circuit without wheel spin or breakout, attempting to achieve a neutral steer. Participant drivers were required to maintain a constant radius and balanced turn, holding Vmax for the remaining two laps. On the final lap, the vehicle was brought to a halt adjacent to the start location as quickly as possible.

## 2.2 Simulator Configuration
The vehicle dynamics in the simulator were provided by a SimPack model of the "XF vehicle" provided by JLR, and had been previously validated against a real vehicle in high friction conditions. The three manoeuvres were implemented as a set of scenarios within the UoLDS environment. A Delft-Tyre model provided support for both non-uniform surface height and friction profiles through the use of two mesh based curve regular grids (CRG). These two grids implemented to a resolution of 50 cm in both X and Y, define values representing the surface height and friction (µ value) at each point on the surface. The CRG surfaces for elevation and friction were built by random sampling from normal distributions representing typical height values (mean: 4mm, standard deviation 1mm) for the lake surface and friction values for both packed snow (mean: 0.4, standard deviation: 0.02) and polished ice (mean: 0.2, standard deviation: 0.005) were used based on braking tests performed on the lake surface (Boer et al., 2015). The visual scene for the Revi test track was modelled using Presagis Creator and rendered within the simulator using the 3D graphics library OpenSceneGraph Figure 2.

Objects representing standard road cones were used to mark out the significant points within each scenario and modelled to allow collisions between any part of the vehicle and a cone to be detected and marked by a small z axis movement in the motion system (to represent the vehicle bumping over the cone) and through a change in the visual state of the cone from vertical to horizontal.

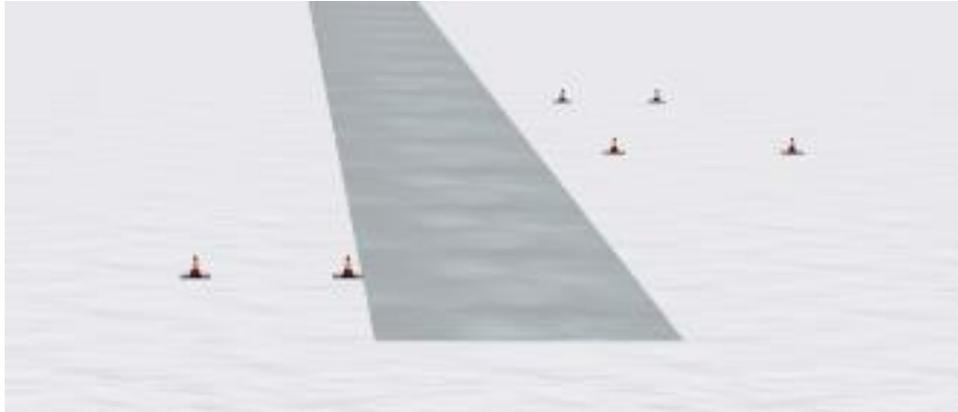

**Figure 2. Simulated Single Lane Change µ Transition**

There were a couple of shortcomings with the implementations of the ice patch in the simulator. The visual representation of the ice surface is wider (at 7 metres) than on the test circuit in Revi (6 metres). This was due to an error in the original specification of the scene. The actual low µ surface (as defined through the CRG file) had the correct width of 6 m, but since the CRG grid was at a resolution of 0.5 m, in practice there was a 0.5 m width on both sides of the patch where friction was linearly interpolated between 0.4 and 0.2.

Another thing to note is that the circular curve was marked with cones in the simulator, not as a ploughed circular lane as in Revi.

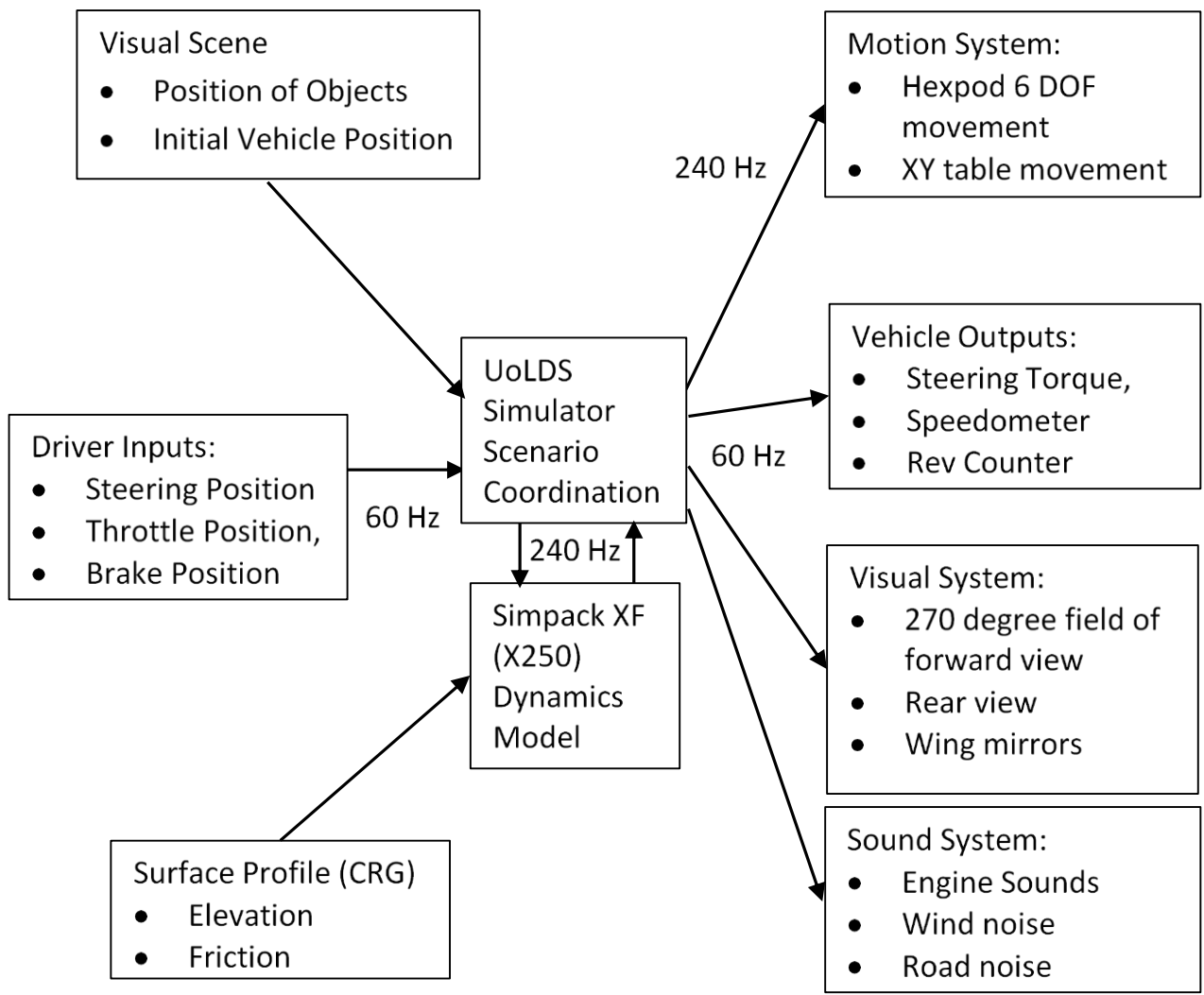

**Figure 3. UoLDS Simulator configuration in the Virtual Sweden study**

The motion system maps the forces supplied by the dynamics model onto an approximation of the movement needed to allow the participant in the vehicle cab to perceive an equivalent force. This is achieved through a 6 degree of freedom hexapod which can supply the heave (+/-0.25 m), roll, pitch and yaw movements (+/-20 degrees) mounted on a 5 metre XY table that can replicate the lateral and longitudinal accelerations of the vehicle. The connection of the various software components and their communication rates are given in Figure 3. The Simpack XF model calculates the steering wheel torque value based on the combination of steering wheel position and trajectory, road wheel angle, vehicle speed, wheel/surface friction and elevation change using the forces generated by the Delft tyre model. The steering system in the simulator can deliver a closed loop torque of up to 8 Nm. Figure 4 shows the "Classical" motion cueing algorithm used.

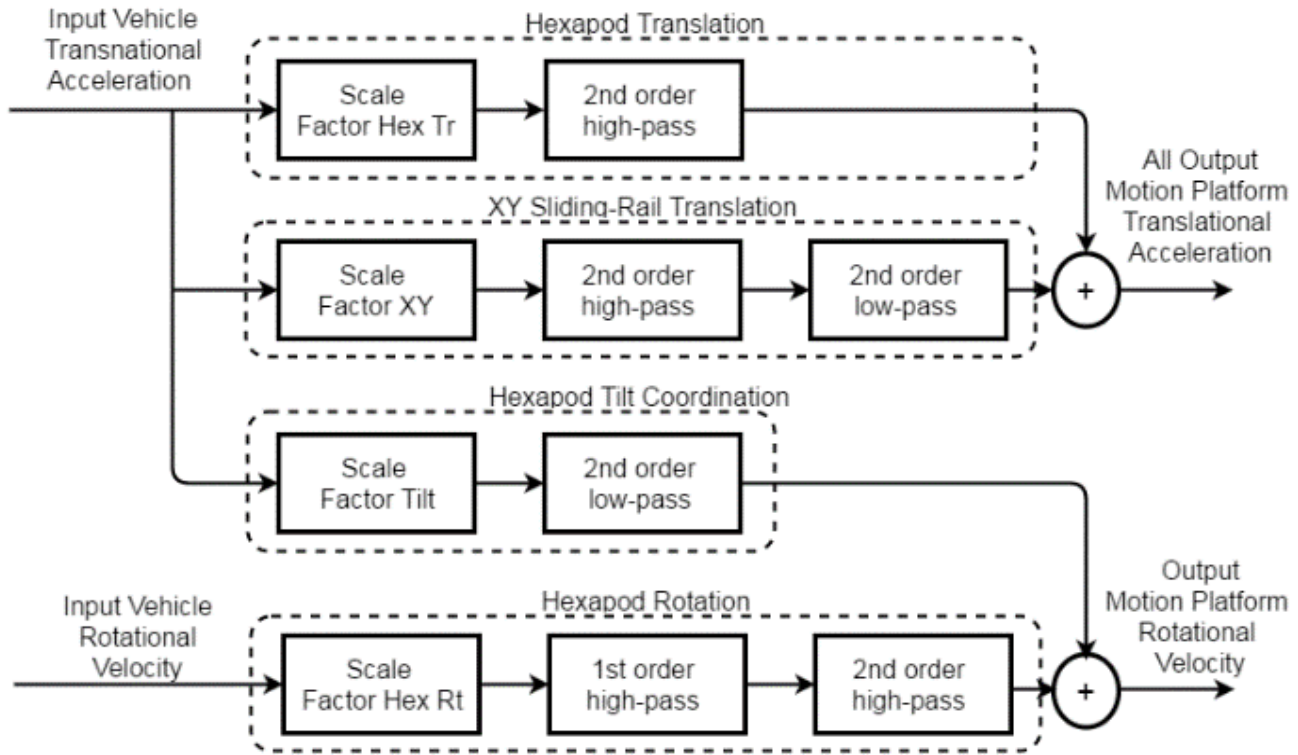

**Figure 4. Schematic illustration of the "classical" algorithm for motion cueing**

## 2.3 Simulator Data Collection

For the data collection in UoLDS, the JLR test drivers came to Leeds in pairs. The first two drivers participated on Dec 9, 2015, and the remaining six over three days in Jan 25-Feb 4, 2016. After arriving in the morning, the drivers were talked through the briefing material that they had been sent beforehand, and they signed a consent form. Next, the experimenter demonstrated the three tasks on a small table top simulator so that the drivers would be familiar with the visual aspect of the tasks already at their first attempts. Furthermore, before starting driving in the Virtual Sweden environment in the UoLDS, the drivers carried out a 10 min familiarisation drive on a simulated UK rural road, to make them accustomed to the simulator vehicle and the simulated motion.

Each driver experienced all nine possible combinations: the three tasks (LCT, SLX, CLV) with the three main simulator configurations (STD, -TRQ, -MOT). In each combination, each driver made four consecutive attempts at the manoeuvre. The order of tasks and simulator configurations was randomised per driver. Before starting a set of four attempts, the drivers were told which task they would be driving, and whether motion and steering torque would be off or on. After each combination of task and simulator configuration, the drivers provided a subjective rating of simulator realism ("*For the task you just drove, how similar would you say that the experience in the simulator is to reality?*") on a visual analogue scale (VAS), i.e. by marking one location on a line, with one end representing 0% and the other 100%. Each driver drove two combinations of task/simulator configuration (i.e. eight task attempts in total), before stopping the simulator to switch drivers.

## 2.4 Real World Data Collection

Data were collected with eight experiment participants in Arjeplog, Sweden, in February-March 2015. The instrumented vehicle (IV) used was a Jaguar XF, equipped with a differential GPS (DGPS) and an inertial measurement unit (IMU) GPS that also enabled estimation of slip angle. Driver control input was logged from the vehicle's CAN network. The XF's Stability Control System (SCS) was turned off since SCS was not available in the SimPack model. Data collection was supervised by the same on-site researcher across all days who ensured the track and any cones were set-up identically over the four day period. Each participant was briefed on the requirements of the study and was required to give informed consent before being allowed a brief period to familiarise themselves with the vehicle and the conditions of the day on the short 1.5 km journey from Track Control to the location of the first task. The tasks were performed by each of the eight participant drivers, with the order counterbalanced by a Latin Square. Each task was carried out three consecutive times by each driver. An ABS braking test on both the polished ice & graded snow surfaces on each day of data collection was performed with the assumption that the maximum acceleration achieved was a measure of the friction available.

# 3. Results

## 3.1 Analysis Method

Statistical hypothesis testing has not been carried out as part of the analyses. Instead, the emphasis has been on effect sizes. The motivation is that statistical significance depends on the size of the experiment and the number of drivers in this case is not suitable. Therefore the main aim is not to test a null hypothesis suggesting that real driving and simulated driving are exactly the same; instead, the aim is to give our best estimates of how *large* the differences between reality and the simulator are, using metrics that are independent of experiment size.

When *d* is reported, this refers to Cohen's *d*, the difference in means divided by the pooled standard deviation. There are many corrections to this estimate of effect size that can be made to adjust for within-subject design and repeated measurements (Lakens, 2013); for simplicity this has not been made. It has been proposed (Cohen, 1988) that $d \approx 0.25$ can be considered a "small" effect, $d \approx 0.5$ a "medium" effect, and $d > 0.8$ a "large" effect.

## 3.2 Data Preparation

The main challenge for the real world was to locate and extract each task attempt separately and to exclude all non-task data from analysis. In the driving simulator it is rather straightforward to find the beginning and end of task attempts, based on vehicle XY position in the simulated world. In the real world the identification of individual task attempts was made based on vehicle speed patterns instead. DGPS data were collected, further enhanced with Kalman filtering together with data from an IMU, and each day the positions of all cones were "marked" with the DGPS receiver to allow analyses that would appreciate position of vehicle relative to the exact task tracks. However, during the subsequent analysis it became apparent that the DGPS positioning had fluctuated somewhat during the day such that the absolute connection between vehicle and cone positions was lost. Therefore an approximate procedure was applied as follows. For each recorded task attempt, a certain "anchor point" within the attempt was located:

LCT: The point where the vehicle had travelled 1 m laterally to the right of the average lateral position in the first second of the task (relative to extraction start point as defined above).

SLX: The point midway between the first leftward lateral position peak (to the left of the second cone) and the first rightward peak (to the right of the third cone)

CLV: An estimated circle centre point, obtained by 500 times selecting at random three points A, B, C from the vehicle trajectory, taking the bisecting normals $N_1$ and $N_2$ of the lines AB and BC, and taking the intersection of $N_1$ and $N_2$ as one circle centre estimate $(X_i, Y_i)$. Then selecting the circle centre as the medians of the $X_i$ and the $Y_i$. The resulting paths in the simulator and the real world for a single driver performing the LCT is given in Figure 5.

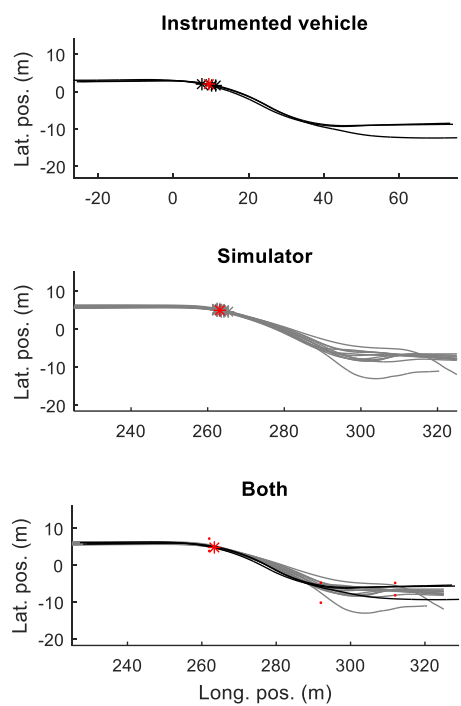

**Figure 5. Resulting Paths Simulator/Real World**

## 3.3 Vehicle Dynamics Fidelity

The dynamical response of the vehicle yaw rate versus steering wheel angle was compare between the simulator and real world for each manoeuvre across all runs and participants. This is shown in Figure 6.

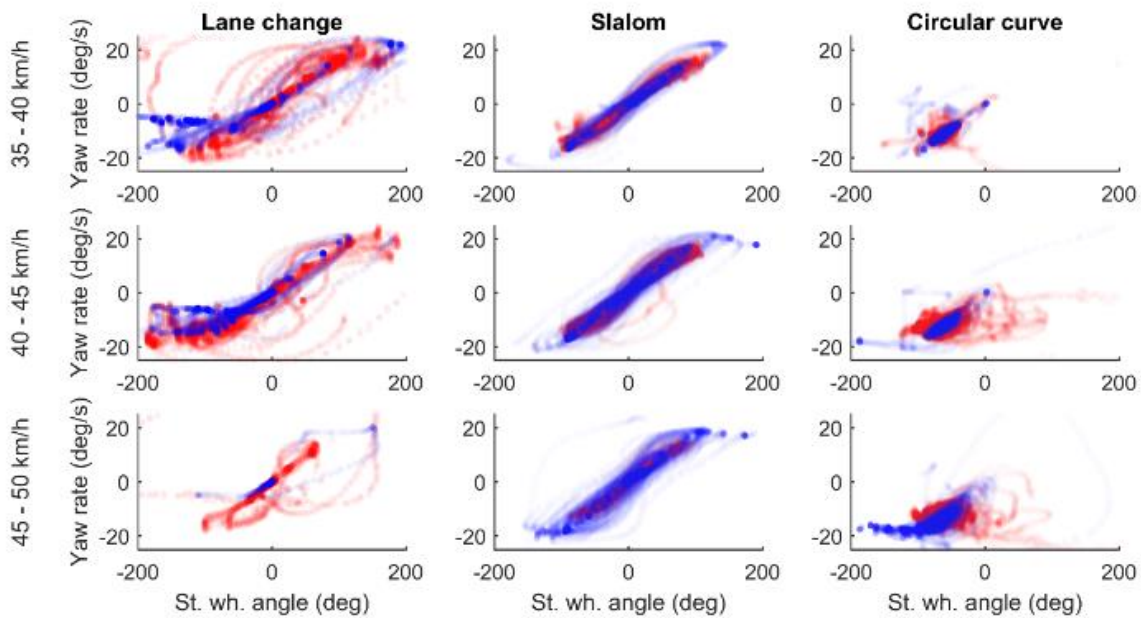

**Figure 6. Yaw rate response in Instrumented Vehicle (Red) and simulator (Blue)**

## 3.4 Subjective Simulator Ratings

Table 2 shows the realism ratings provided by drivers for the three tasks, across all tested simulator configurations, as well as the Cohen d compared to the standard configuration. It should be noted that the three simulator configurations were rank ordered the same in all three tasks: STD, -TRQ, -MOT in descending order of realism. Large effects are highlighted. For the LCT and CLV manoeuvres turning off the motion had a large effect on driver reported realism. For the SLX turning off either torque or motion had a large effect.

**Table 2. Cohen d and Rating for Realism**

| Measure | STD | -TRQ | -MOT |
|---|---|---|---|
| LCT Avg. Real | 70% | 60% | 40% |
| LCT Cohen d |  | -0.51 | **-1.92** |
| SLX Avg. Real | 72% | 60% | 40% |
| SLX Cohen d |  | **-0.81** | **-1.54** |
| CLV Avg. Real | 65% | 60% | 42% |
| CLV Cohen d |  | -0.35 | **-0.98** |

## 3.5 Objective Results

When fitting driver model parameters to data, a brute force grid search was used for all nonlinear parameters (e.g. $T_R$), and for each tested combination of these nonlinear parameters, all linear parameters (e.g. $K$) were estimated using least squares fitting. The DPYRE driver model yielded a goodness of fit ($R^2$, or variance explained) on average for the LCT above 0.6 and SLX on average being above 0.8. For the CLV, the original DPYRE fit the data very poorly (roughly 0.2) and therefore the intermittent DPYRE was explored. The goodness of fit improved substantially but is still less than the other two manoeuvres. Tables 3, 4, 5 give the Cohen d for each measure of the Utility Triplet compared to the real world. Those elements that had a large effect are highlighted in bold.

**Table 3. Cohen d for LCT Manoeuvre**

| Measure | STD | -TRQ | -MOT |
|---|---|---|---|
| Cones Hit | **0.87** | **0.82** | **0.94** |
| Speed Var. | **1.20** | **0.92** | 0.77 |
| Max Lat Accel | 0.60 | 0.50 | 0.21 |
| Initial Speed | -0.73 | -0.57 | -0.66 |
| Total Lat Travel | 0.29 | 0.49 | 0.53 |
| SWRR 1 deg | 0.55 | 0.37 | 0.31 |
| SWRR 10 deg | 0.00 | -0.01 | 0.01 |
| Tr | -0.09 | -0.01 | 0.02 |
| K | 0.58 | 0.27 | **1.15** |
| RMS | 0.06 | 0.18 | -0.24 |

For the LCT, all simulator configurations had a large effect on the average number of cones hit with the real vehicle being close to zero and in the simulator an average of 0.5 cones were hit per trial across all configurations. There was also a large effect on speed variation for the STD and –TRQ configurations with the speed variability at 3 km/h in the real vehicle and 5.5 km/h in the STD and –TRQ configurations. Finally in the –MOT configuration there was a large effect on the driver model parameter K which increased from 13 in the real world to 18 in the simulator.

**Table 4. Cohen d for SLX Manoeuvre**

| Measure | STD | -TRQ | -MOT |
|---|---|---|---|
| Cones Hit | -0.20 | 0.16 | 0.43 |
| Speed Var. | 0.35 | 0.19 | 0.21 |
| Average Speed | 0.60 | 0.50 | 0.21 |
| Lateral Amp | 0.72 | **1.04** | 0.51 |
| Lat Amp Var | -0.16 | 0.24 | 0.04 |
| SWRR 1 deg | 0.69 | 0.72 | **1.41** |
| SWRR 10 deg | 0.77 | **0.91** | **1.44** |
| Tr | -0.79 | -0.30 | -0.69 |
| K | -0.18 | 0.76 | -0.55 |
| RMS | **0.82** | **1.17** | **1.35** |

In the SLX, the lateral amplitude taken in the manoeuvre increased from 2.8 metres in the real world to 3.6 metres in the –TRQ case. The –MOT case had a large effect on the SWRR 1 deg increasing the rate from 0.55 Hz in the real world to 0.8 Hz. Both the –TRQ and –MOT configurations had a large effect on the SWRR 10 deg increasing the rate from 0.5 Hz in the real world to 0.6 for the –TRQ case and 0.7 Hz for the –MOT case. The simulator had a large effect on the RMS driver model error with the error increasing from 0.55 in the real world to 1 across the simulator configurations.

**Table 5. Cohen d for CLV Manoeuvre**

| Measure | STD | -TRQ | -MOT |
|---|---|---|---|
| Speed Var. | 0.54 | 0.47 | 0.75 |
| Radius Var. | 0.53 | 0.52 | 0.48 |
| Mean Speed | **1.13** | **1.25** | **1.20** |
| Mean Radius | **-1.97** | **-1.98** | **-1.78** |
| SWRR 1 deg | **-0.96** | **-0.81** | **-0.81** |
| SWRR 10 deg | 0.01 | 0.22 | 0.11 |
| Tr | -0.32 | 0.12 | 0.02 |
| K | 0.30 | 0.46 | 0.39 |
| RMS | 0.32 | 0.40 | 0.33 |
| Average Adj. | 0.24 | 0.27 | 0.38 |

Finally for the CLV case all the simulator configurations had a large effect on three measures with the mean speed increasing from 44 km/h in the real world to 49 km/h in the simulator. The mean radius decreased from 55.5 metres in the real world to 53.5 metres in the simulator. Finally the SWRR 1 deg decreased from 1 Hz in the real world to 0.8 Hz across the simulator configurations.

## 4. Discussion

To answer the first research question: "In the highest-specification UoLDS simulator configuration, how similar is driver behaviour in the winter testing tasks to behaviour in a real vehicle?", the simulator seems to have performed very well for the SLX manoeuvre. For the LCT and CLV manoeuvres there are large effects from using the simulator. Reviewing Figure 6, in the SLX the real world and simulator yaw rates matched much more closely than the LCT and CLV. In the LCT the simulator, shows a clearly bimodal distribution of yaw rates at high negative (rightward) steering wheel angles. This bimodality is due to the difference in friction between snow and ice patch, but this did not seem to occur in reality. Based on this observation, an in depth analysis for limit handling of the effective friction coefficient in the real world for the ice and snow cases showed that the ice was typically a higher friction limit than used in the simulator while the snow had a lower friction limit in the real world than in the simulator. It is hypothesized that the lower ice friction limit in the simulator led to a greater number of cones hit in the LCT manoeuvre while the high snow friction limit in the simulator led to a faster drive in the CLV manoeuvre. The higher snow friction limit may have also led to a lower SWRR 1 deg in the CLV manoeuvre for the simulator since the task in the simulator may have been easier. This is in line with research from Gómez et al. (Gil Gómez et al., 2017) in which reduction of the available friction led to less repetitive results and that changes in the real world surface condition could cause a large spread of the data. The decrease in mean radius in the CLV manoeuvre could be due to the higher friction level or possibly due to perceptual differences between the simulator and the real world since the driver was asked to drive half a car width away from the cones. The dome of the UoLDS has a radius of 2 m and the vehicle was right hand drive with a clockwise direction around the cones. This leaves a gap between the right hand side of the vehicle and the screen in which there is no additional visual information because the visual display does not project onto the floor of the dome. This gap may have made it more difficult for the driver to judge their position relative to the cones.

For the second research question: "How is behavioural fidelity in the winter testing tasks affected by altering the vestibular and haptic cues that are available to the drivers?", For the LCT task, behavioural fidelity was roughly the same for –TRQ as for STD, whereas when moving to –MOT, the driver model gain K started deviating. For SLX, –TRQ and –MOT affected several of the time series parameters. For CLV, there did not seem to be any identifiable difference between the simulator configurations.

For the third research question: "How do the drivers' subjective impressions of simulator realism relate to objective measures of behavioural fidelity (subjective impressions)?", the general pattern across the Utility Triplet metrics was that objective behavioural fidelity deteriorated somewhat when moving from STD to –TRQ, and even more when moving from STD to –MOT. This aligns well with the subjective realism ratings, which placed these three configurations in this same order. The subjective realism was consistent across manoeuvres but the objective data were more manoeuvre dependent.

# 5. Conclusion

The results of this first full application of the Utility Triplet methodology suggest that the developed "Virtual Sweden" environment achieved rather good behavioural fidelity in the sense of preserving absolute levels of many measures of behaviour. The slalom task came across as especially well replicated in the simulator. In the lane change and circular curve tasks, there were some specific notable differences in behaviour between reality and simulator. These seem to be due to imperfections in the simulated tyre-ground interactions near the limit of stability. Overall, the obtained results can be taken to suggest that if the tyre-ground interaction limitations can be addressed, high-fidelity behaviour should be achievable in the simulator across all the studied tasks.

The objective measures of behaviour were complemented with subjective ratings of simulator realism. The preferred simulator configurations were on average rated by drivers at about 70 % realism (with 100 % corresponding to real-vehicle driving).

Besides driving the UoLDS at full capacity, the drivers also experienced various scaled-down versions of the simulator; removing either the steering torque feedback or the motion feedback altogether. The utility triplet analyses were able to pick out a range of effects of these manipulations on driver behaviour, largely in line with predictions based on existing literature, and interestingly also aligning well with the subjective realism: When perceptual cues were removed, and especially so when removing motion cues altogether, the general patterns were that behavioural fidelity deteriorated, steering efforts increased, and steering gains of fitted driver models increased.

Finally the conclusions have been drawn on an effects size analysis and additional data should be collected to ensure that there is statistical evidence to support them.

# 6. Acknowledgments

This work was supported by Jaguar Land Rover and the UK-EPSRC Grant EP/K014145/1 as part of the jointly funded Programme for Simulation Innovation (PSi). Upon publication of this paper, the research data supporting it will be made publicly available.